\documentstyle[graphicx,12pt]{article}

\begin{document}
\begin{titlepage}
\vspace{5cm}
\title{\bf Inflation in Gauss--Bonnet Brane Cosmology}
\vspace{3cm}
\author{James E. Lidsey$^1$ and N. J. Nunes$^2$ \vspace{1cm} \\ 
\normalsize{\em  Astronomy Unit, School of Mathematical Sciences,}  \\ 
\normalsize{\em Queen Mary, University of London, Mile End 
Road, LONDON, E1 4NS, U.K.} }
\date{\today}
\maketitle
%\vspace{1.5cm}
\begin{abstract}
\noindent
The effect of including a Gauss--Bonnet contribution in the bulk action
is investigated within the context of
the steep inflationary scenario. When inflation is driven by an exponential
inflaton field, this Gauss--Bonnet term allows the spectral
index of the scalar perturbation spectrum to take values
in the range 0.944 and 0.989, thereby
bringing the scenario in closer agreement with the most recent
observations. Once the perturbation spectrum is normalized to the
microwave background temperature anisotropies, the value of the
spectral index is determined by the Gauss--Bonnet
coupling parameter and the tension of the brane and
is independent of the logarithmic slope of the potential.
\end{abstract}
\thispagestyle{empty}

\vspace{1.5cm} 

\vspace{1cm}
\noindent $^1$Email: J.E.Lidsey@qmul.ac.uk \\
\noindent $^2$Email: N.J.Nunes@qmul.ac.uk
 
%\vskip-18cm
%\rightline{hep-ph/0303xxx}
%\vskip3in
\end{titlepage}
%%%%%%%%%%%%%%%%%%%%%%%%%%%%%%%%%%%%%%%%%%%%%%%%%%%%%%%%%%%%%%%%%%%%

\newpage

%%%%%%%%%%%%%%%%%%%%%%%%%%%%%%%%%%%%%%%%%%%%%%%%%%%%%%%%%%%%%
\section{\label{sec1}Introduction}
%%%%%%%%%%%%%%%%%%%%%%%%%%%%%%%%%%%%%%%%%%%%%%%%%%%%%%%%%%%%%
%\setcounter{equation}{0}
%\def\theequation{\thesection.\arabic{equation}}

Recent high precision measurements by the Wilkinson Microwave
Anisotropy Probe (WMAP) of acoustic peak structure 
in the anisotropy power spectrum of the 
cosmic microwave background (CMB) radiation 
have provided strong evidence that 
the universe is very close to critical density
and that large--scale structure developed through 
gravitational instability from a primordial spectrum of 
adiabatic, Gaussian and nearly scale--invariant density perturbations
\cite{wmap,bridle}. 
These observations are consistent with 
the cornerstone predictions of the simplest class of 
inflationary models \cite{simplest,perturbations}. 
(For a review, see, e.g., Ref.~\cite{lidlyth}). 

In view of these developments, it is important to further our understanding of
the inflationary scenario from  
a theoretical perspective. Presently, 
there is considerable interest in inflationary models 
motivated by superstring and M--theory \cite{string,lwc}. 
In particular, much attention has focused on the 
braneworld scenario,  
where our observable, four--dimensional universe is regarded as a 
domain wall (co--dimension 1 brane) embedded in a higher--dimensional
bulk space \cite{early,RSII}. An important realisation of this picture is 
provided by the 
Randall--Sundrum type II scenario (RSII), where a spatially isotropic and 
homogeneous brane propagates in a five--dimensional 
Schwarzschild--Anti--de Sitter (AdS) space \cite{RSII}. Even though the 
fifth dimension is non--compact, the graviton zero--mode is localized  
on the brane due to the non--factorizable geometry of the higher--dimensional 
space. Moreover, it has been shown 
within the context of the AdS/CFT correspondence \cite{adscft} that 
the RSII model is equivalent to four--dimensional 
gravity coupled to a conformal field theory (CFT) \cite{hrr}. 
In this interpretation, the Einstein--Hilbert 
action on the boundary of the AdS space arises as a surface counter term that 
is introduced to cancel the divergences arising in the five--dimensional 
gravitational action. 

One approach to developing the braneworld scenario in a 
more string theoretic setting is to include higher--order 
curvature invariants in the bulk action
\cite{HD,GBrefs,onlylocal,gbf,davis,gw,bcdd}. 
Such terms arise in the AdS/CFT correspondence as next--to--leading order 
corrections in the $1/N$ expansion of the CFT \cite{largeN}. 
More specifically, the Gauss--Bonnet combination
arises as the leading order quantum correction in the 
heterotic string effective action and in five dimensions 
represents the 
unique combination of curvature invariants 
that leads to second--order field equations 
in the metric tensor \cite{stringGB,bd,d86}. It has been further shown that 
localization of 
the graviton zero--mode on the brane is possible when 
such a term is included in the bulk action \cite{onlylocal}.  

An investigation into the effects of a Gauss--Bonnet term 
on inflationary braneworld models in the RSII scenario  
is therefore well motivated and this is the purpose of the 
present paper. We develop a model of inflation known as {\em steep
inflation}, where potentials that are too steep to support inflation 
in standard cosmology are able to drive a period of inflationary 
expansion due to corrections in the Friedmann equation that arise 
as a consequence of the brane dynamics 
\cite{steep,steep1,steepgw,steeppbh,steepcurv,maartens,nunes,othersteep}. 
A related model has been considered recently within the 
context of the `two measures' theory, where an integration 
measure that is independent of the metric tensor is introduced into the 
action \cite{guendelman,gk}. 
By imposing global scale invariance on such a theory, it is found 
that the potentials of the scalar field are restricted to be of 
an exponential form. Inflation is then possible in this 
theory due to the additional 
friction terms that are present in the cosmological field equations. 
The implications for inflationary 
cosmology of introducing higher--order curvature invariants 
into the theory have also been investigated \cite{gk}. 

One
of the key predictions of the simplest braneworld models 
of steep inflation is that the spectral index of the 
scalar perturbation spectrum should deviate significantly 
from unity, $n_S \approx 0.93 
- 0.94$ \cite{steep,steep1}, and such values are presently 
disfavoured by recent observations \cite{wmap,bridle}. However, 
for the case of an exponential inflaton potential, we find
that 
the inclusion of a Gauss--Bonnet term in the bulk action 
can result in spectra where $n_S \approx 1$.

%%%%%%%%%%%%%%%%%%%%%%%%%%%%%%%%%%%%%%%%%%%%%%%%%%%%%%%%%
\section{Friedmann equation}
%%%%%%%%%%%%%%%%%%%%%%%%%%%%%%%%%%%%%%%%%%%%%%%%%%%%%%%%%
%\setcounter{equation}{0}
%\def\theequation{\thesection.\arabic{equation}}

The five--dimensional bulk action for the Gauss--Bonnet braneworld 
scenario is given by 
\begin{equation}
\label{5daction}
S_{\cal{M}} = \frac{1}{2\kappa^2_5} \int_{\cal{M}} d^5x \sqrt{-g} 
\left[ R -2\Lambda +\alpha \left( R^2 -4R_{ab}R^{ab} 
+R_{abcd}R^{abcd} \right) \right] +S_{\partial \cal{M}} +S_{\rm mat} 
\,,
\end{equation}
where $\alpha >0$ represents the Gauss--Bonnet coupling, 
$\Lambda < 0$ is the bulk cosmological constant and 
$\kappa_5^2= 8 \pi M_5^{-3}$ determines the five--dimensional 
Planck scale. The full action also includes the appropriate boundary 
term, $S_{\partial {\cal{M}}}$, required to cancel normal derivatives
of the 
metric tensor 
that arise when varying the action with respect to the metric \cite{myers}. 
Matter on the brane is incorporated by including a term of the 
form $S_{\rm mat} = \int_{\partial {\cal{M}}} d^4 x\sqrt{-h} 
{\cal{L}}_{\rm mat}$, where $h$ is the induced metric on the brane
and ${\cal{L}}_{\rm mat}$ is the matter lagrangian. 

Cosmological dynamics on the brane arises due to its
motion through the static bulk space \cite{kraus}. 
The bulk field equations admit AdS space as a solution \cite{bd,cai}
and, in this case, the induced metric on the brane corresponds to the 
spatially isotropic and homogeneous Friedmann--Robertson--Walker
(FRW) line--element, where the scale factor, $a(t)$, is related
to the position of the brane in the bulk. 
In this paper, we neglect the effects 
of spatial curvature on the brane, 
since we are interested in inflationary cosmology 
and such terms are rapidly redshifted away.
The effective Friedmann equation for our universe may be derived 
from a generalization of Birkhoff's theorem \cite{gbf}. 
A more geometrical approach may be taken \cite{davis} by varying the 
boundary term or employing the formalism of differential forms \cite{gw}. 
Imposing a 
${\bf  Z}_2$ symmetry across the brane and assuming that 
a perfect fluid matter source is confined to the brane then 
results in a Friedmann equation of the form \cite{gbf,davis,gw}
\begin{equation}
\label{friedmann}
H^2= \frac{c_+ +c_- -2}{8\alpha} \,,
\end{equation}
where
\begin{equation}
\label{defc}
c_{\pm} = \left\{ \left[ \left( 1+\frac{4}{3}\alpha \Lambda \right)^{3/2} 
+ \frac{\alpha}{2} \kappa_5^4 \sigma^2 \right]^{1/2} 
\pm \sqrt{\frac{\alpha}{2}} \kappa_5^2 \sigma \right\}^{2/3} \,,
\end{equation}
and $\sigma$ represents the energy density of the matter sources. 

Conservation of energy--momentum of the matter on the brane 
follows directly from the Gauss--Codazzi equations. 
For a perfect fluid matter source, these reduce to the familiar 
form
\begin{equation}
\label{conservation}
\dot{\sigma}+3H \left( \sigma +p \right) = 0 \,,
\end{equation}
where $p$ represents the pressure of the 
fluid and a dot denotes differentiation with respect to 
synchronous time on the brane. 

Eqs.~(\ref{friedmann}) and (\ref{conservation}) 
are sufficient to fully determine the cosmic dynamics 
on the brane once an equation of state has been specified 
for the matter sources. 
Such an analysis can be simplified considerably by 
defining a new variable, $x$: 
\begin{equation}
\label{defx}
\sigma \equiv \left( \frac{2b}{\alpha \kappa^4_5} \right)^{1/2}
\sinh x \,,
\end{equation}
and a new constant, $b$: 
\begin{equation}
\label{defb}
b \equiv \left( 1+\frac{4}{3} \alpha \Lambda \right)^{3/2} \,.
\end{equation}
It then follows from Eq.~(\ref{defc}) that 
$c_{\pm} = b^{1/3} \exp(\pm 2x/3) $
and this implies that 
the Friedmann equation (\ref{friedmann}) reduces to the particularly 
simple form
\begin{equation}
\label{friedmannx}
H^2 = \frac{1}{4\alpha} \left[ b^{1/3} \cosh \left( \frac{2x}{3} \right) 
-1 \right] \,.
\end{equation}

Although the bulk action contains three parameters, $\{ \kappa_5 , \Lambda , 
\alpha \}$, 
the standard form of the Friedmann equation must be recovered at low energies
and this constraint implies that the parameters are not 
independent. In particular, 
substituting Eq.~(\ref{defx}) into Eq.~(\ref{friedmannx}) and expanding to 
quadratic order in the energy density implies that 
\begin{equation}
\label{friedmannquad}
H^2 = \frac{\kappa_5^4}{36b^{2/3}} \sigma^2 + 
\frac{1}{4\alpha}(b^{1/3} -1) \,.
\end{equation}
We now invoke the standard assumption that the energy density on the brane 
can be separated into two
contributions, the ordinary matter component, $\rho$, and the brane
tension, $\lambda >0$, such that $\sigma  = \rho  +\lambda$. We then
obtain the modified Friedmann equation 
\begin{equation}
\label{FriedmannRSII}
H^2 = \frac{\kappa_4^2}{3} \rho \left[ 1+\frac{\rho}{2\lambda} \right]
+ \frac{\Lambda_4}{3} \,,
\end{equation}
where the four--dimensional cosmological constant is defined as
\begin{equation}
\Lambda_4 = \frac{3}{4\alpha}(b^{1/3} -1) + 
\frac{\kappa_5^4}{12b^{2/3}} \lambda^2 \, .
\end{equation}
The standard form of the Friedmann equation is recovered at sufficiently 
low energy scales $(\rho \ll \lambda )$ by identifying 
\begin{equation}
\label{kappas}
\kappa_4^2 \equiv \frac{8\pi}{m_{\rm Pl}^{2}} = \frac{\kappa_5^4
  \lambda}{6b^{2/3}} \, ,
\end{equation}
where $m_{\rm Pl}$ is the four--dimensional Planck scale. Finally, 
the four--dimensional cosmological constant vanishes when the brane tension 
satisfies 
\begin{equation}
\label{tensiontune1}
\lambda = \frac{3}{2} (1-b^{1/3}) \frac{1}{\alpha \kappa_4^2} \,.
\end{equation}
It is straightforward to verify 
that in the limit of $\alpha \rightarrow 0$, Eq.~(\ref{tensiontune1}) 
reduces to the RSII constraint such that  $\lambda = -\Lambda/\kappa_4^2$
\cite{RSf}.

This concludes our discussion on the parameters of the model. 
In the following Section, we consider the dynamics of inflationary cosmology 
within the context of the Friedmann equation (\ref{friedmannx}). 

%%%%%%%%%%%%%%%%%%%%%%%%%%%%%%%%%%%%%%%%%%%%%%%%%%%%%%%%%%%%
\section{Steep inflation}
%%%%%%%%%%%%%%%%%%%%%%%%%%%%%%%%%%%%%%%%%%%%%%%%%%%%%%%%%%%%%%%
%\setcounter{equation}{0}
%\def\theequation{\thesection.\arabic{equation}}

%%%%%%%%%%%%%%%%%%%%%%%%%%%%%%%%%%%%%%%%%%%%%%%%%%%%%%%%%%%%%%%%
\subsection{Conditions for inflation}
%%%%%%%%%%%%%%%%%%%%%%%%%%%%%%%%%%%%%%%%%%%%%%%%%%%%%%%%%%%%%%%%

For a general equation of state,  
$p = \left[ \gamma (x) -1 \right] \sigma $,
where $\gamma (x)$ is an arbitrary function, 
the condition for inflation, $\dot{H}+H^2 >0$, becomes
\begin{equation}
\label{inflation}
\cosh ( 2x/3 ) -\gamma (x) 
\tanh (x) \sinh ( 2x/3) > b^{-1/3} \,.
\end{equation}
The origin of the $b^{-1/3}$ term on the right hand side of inequality 
(\ref{inflation}) arises directly from the effective 
negative cosmological constant term, $-1/(4 \alpha )$, in the 
Friedmann equation (\ref{friedmann}). 
In the high--energy limit, $x \gg 1$, condition (\ref{inflation}) 
reduces to 
\begin{equation}
( 1-\gamma ) e^{2x/3} > \frac{2}{b^{1/3}} \,,
\end{equation}
or equivalently, $\gamma < 1$.
In the corresponding limit, $x \ll 1$, 
condition (\ref{inflation}) simplifies to 
\begin{equation}
\label{inflationlow1}
2(1-3\gamma) x^2> 9 \left( b^{-1/3} -1 \right) \,,
\end{equation}
or equivalently, $\gamma < 1/3$.
Eq.~(\ref{inflationlow1}) 
implies that a necessary condition for inflation is that 
$p <-2\sigma/3$. In the limit where the energy density of 
the matter dominates the brane tension, this is equivalent 
to the condition for inflation to proceed in the standard 
RSII scenario \cite{maartens}. Consequently, the Gauss--Bonnet 
contribution does not alter the condition for 
inflation to end when $x$ is small.  

%%%%%%%%%%%%%%%%%%%%%%%%%%%%%%%%%%%%%%%%%%%%%%%%%%%%%%%%%%
\subsection{Inflationary dynamics}
%%%%%%%%%%%%%%%%%%%%%%%%%%%%%%%%%%%%%%%%%%%%%%%%%%%%%%%%%

We assume that during inflation, our braneworld is 
dominated by a single, minimally coupled scalar field, $\phi$, that is  
confined to the brane and self--interacts through a potential, 
$V(\phi )$. The conservation equation (\ref{conservation}) then
implies that 
\begin{equation}
\label{scalareom}
\ddot{\phi} +3H\dot{\phi} +V'=0 \,,
\end{equation}
where a prime denotes differentiation with respect to the scalar 
field.

In conventional cosmology, the potential must be 
sufficiently flat for the universe to undergo a phase of 
accelerated expansion. The key feature of the steep inflationary scenario is 
that the quadratic corrections to the Friedmann equation arising in the RSII 
scenario enhance the friction acting on the scalar field 
as it rolls down its potential, thereby enabling a steeper class of potentials 
to support inflation \cite{steep,maartens}. 
Generically, steep inflation proceeds in the region of 
parameter space where 
$\sigma \approx \rho \gg \lambda$ and naturally comes to an end when 
$\rho \approx \lambda$, since the conventional cosmological 
dynamics is then recovered. 

We therefore focus our attention on the 
region of parameter space where the inflaton potential dominates 
the brane tension and further 
assume the slow--roll approximation, $\dot{\phi}^2 \ll V$ 
and $|\ddot{\phi}| \ll H |\dot{\phi}|$. Hence, with $\rho \approx V$
and employing Eqs.~(\ref{defx}) and (\ref{kappas}) we write
\begin{equation}
\label{Vsinhx}
V = \left(\frac{\lambda b^{1/3}}{3 \alpha \kappa_4^2} \right)^{1/2} 
\sinh x \,.
\end{equation}
The slow--roll 
parameters, $\epsilon \equiv -\dot{H}/H^2$,
$\eta \equiv V''/(3H^2)$ and $\xi^2 \equiv V''' V'/(3 H^2)^2$, 
may then be written 
in the form 
\begin{eqnarray}
\label{epsilon}
\epsilon &=& \left( \frac{2\lambda}{\kappa_4^2}~\frac{V'^2}{V^3} \right) 
\left[ \frac{2b^{2/3}}{27}~ 
\frac{\sinh (2x/3) \, \tanh x \, \sinh^2 x}{\left[ 
b^{1/3} \cosh (2x/3) -1 \right]^2} \right] \,, \\
\label{eta}
\eta &=& \left( \frac{2\lambda}{\kappa_4^2}~ \frac{V''}{V^2} \right) 
\left[ \frac{2b^{1/3}}{9}~ 
\frac{\sinh^2 x}{b^{1/3} \cosh (2x/3) -1} \right] \,, \\
\label{xi}
\xi^2 &=& \left( \frac{4 \lambda^2}{\kappa_4^4} ~\frac{V'''V'}{V^4}
\right) \left[ \frac{4 b^{2/3}}{81} ~\frac{\sinh^4 x}{\left[ 
b^{1/3} \cosh (2x/3) -1 \right]^2} \right] \,,
\end{eqnarray}
and inflation occurs for $\epsilon < 1$. 
The terms in the curved brackets represent the slow--roll 
expressions for the RSII scenario \cite{maartens,nunes} and, consequently,  
the terms in the square brackets may be viewed as the modifications 
to the RSII inflationary scenario due to the Gauss--Bonnet 
contribution. These modifications to the 
slow--roll parameters are
monotonically decreasing functions of $x$ and tend 
to unity from above for $x \ll 1$ and $\alpha \rightarrow 0 $. 
This implies that for a given 
potential, the introduction of a Gauss--Bonnet term into the bulk action 
tightens the condition for slow--roll inflation 
relative to the corresponding condition for the RSII 
scenario. This does not indicate that steep inflation 
can not proceed, however, since the overall effect of the extra contributions 
to the Hubble parameter is to introduce additional 
friction into the scalar field dynamics. 

The number of e--folds of inflationary expansion, $N \equiv 
\int Hdt$, is deduced in terms of the variable, $x$, 
by employing Eqs.~(\ref{friedmannx}) and (\ref{Vsinhx}):
\begin{equation}
\label{Ngeneral}
N (x) = -\left( \frac{27}{16 b^{1/3}} 
~\frac{\kappa_4^2}{\alpha \lambda}
\right)^{1/2} \int_{x_N}^{x_{\rm end}} dx \left( \frac{d\phi}{dx} \right)^2 
\frac{b^{1/3} \cosh (2x/3) -1}{\cosh x} \,,
\end{equation}
where $x_{\rm end}$ denotes the value of $x$ when inflation ends.

%%%%%%%%%%%%%%%%%%%%%%%%%%%%%%%%%%%%%%%%%%%%%%%%%%%%%%%%%
\subsection{Density perturbations}
%%%%%%%%%%%%%%%%%%%%%%%%%%%%%%%%%%%%%%%%%%%%%%%%%%%%%%%%%

The perturbations generated quantum mechanically from a single inflaton field
during inflation are adiabatic \cite{perturbations}. 
The curvature perturbation 
on a uniform density hypersurface is conserved on large scales  
as a direct result of energy--momentum conservation
on the brane \cite{wands}. 
This implies that the amplitude of a given mode when  
re--entering the Hubble radius after inflation is given by 
$A_S^2 = H^4/ (25\pi^2 \dot{\phi}^2)$, where we have adopted the normalization 
conventions of Ref.~\cite{llkcba} and the right--hand side of this expression 
is evaluated when the mode first goes beyond the Hubble radius
during inflation, i.e., when the comoving 
wavenumber, $k$, satisfies $k=aH$.  
Substituting Eqs.~(\ref{friedmannx}) and (\ref{scalareom}) implies that 
\begin{equation}
\label{scalar}
A_S^2= \left. 
\left( \frac{1}{600\pi^2}~ \frac{\kappa_4^6V^6}{\lambda^3 V'^3} \right)
\left[ \frac{729}{8b}~ 
\frac{\left[ b^{1/3} \cosh (2x/3) -1 \right]^3}{\sinh^6 x} \right]
\right|_{k=aH} \,,
\end{equation}
where, as before, the quantity in the square 
bracket represents the Gauss--Bonnet 
correction to the RSII result \cite{maartens}. 
The COBE normalization is $A_S^2 =4 \times 10^{-10}$ \cite{norm}. 
It can be shown from 
Eqs.~(\ref{Vsinhx}), (\ref{epsilon}), (\ref{eta}) and (\ref{xi}) that
the spectral index of the scalar spectrum, $n_S-1 \equiv 
d\ln A_S^2/d \ln k$, is given by
\begin{equation}
\label{scalartilt}
n_S = 1 -6\epsilon +2 \eta \,,
\end{equation}
and its running has the form
\begin{equation}
\label{runns}
\frac{d n_S}{d \ln k} = 6(\zeta-4) \epsilon^2 + 16 \epsilon \eta -
2 \xi^2 \,,
\end{equation}
where we have defined the function $\zeta(x)$ (not a slow--roll
parameter) as
\begin{equation}
\zeta(x) \equiv  \frac{3}{b^{1/3}} \left[b^{1/3} \cosh(2x/3)-1\right] \left[
\frac{2}{3}~\frac{\cosh(2x/3)}{\sinh^2(2x/3)} - \frac{\sinh x}{\cosh x
  \sinh(2x/3)} \right] \,.
\end{equation}
This quantity appears as a consequence of the Gauss--Bonnet contribution and
approaches unity for $x \ll 1$ and $\alpha \rightarrow 0$. 
Hence, we recover the RSII expression
for the running of the spectral index in this limit \cite{nunes}.

At sufficiently high energies $(x \gg 1)$,  the dependence of the Friedmann 
equation (\ref{friedmannx}) on the density of matter is of an 
unconventional 
form, $H^2 \propto \rho^{2/3}$. In light of this, it is instructive 
to first consider inflation based on a generic Friedmann 
equation $H^2=A\rho^q$, where $\{ A,q \}$ are 
arbitrary constants. In this case, the slow--roll parameters 
are
\begin{equation}
\label{epsilon2/3}
\epsilon = \frac{q}{6A} \frac{V'^2}{V^{1+q}} \,, \qquad 
\eta = \frac{1}{3A} \frac{V''}{V^{q}} \,,
\end{equation}
whereas the amplitude of density perturbations 
is given by $A^2_S = (9A^3V^{3q})/(25\pi^2 V'^2)$. The 
scalar spectral index may then be evaluated: 
\begin{equation}
\label{scaleinv}
n_S-1 = \frac{1}{AV^{q-1}} \left( \frac{2V''}{3V}
-\frac{qV'^2}{V^2} \right) \,.
\end{equation}

The functional form of the inflaton potential
that results in a precisely scale--invariant spectrum, $n_S=1$, may now 
be deduced. Eq.~(\ref{scaleinv}) 
reduces to a differential equation 
in the inflaton potential, $2VV'' = 3qV'^2$, and, by employing the 
identity $2V''=dV'^2/dV$, this condition can be solved in full 
generality. For $q\ne 2/3$, we find that the potential 
has a power law form, $V \propto 
\phi^{2/(2-3q)}$. On the other hand, for the case of interest in this paper, 
$q=2/3$, the potential has a purely exponential form and can be written as 
\begin{equation}
\label{exppot}
V=V_0 e^{\beta \kappa_4 \phi} \,,
\end{equation}
where  $V_0 $ is an arbitrary constant and $\beta$ 
determines the self--coupling of the field. 

The condition for the 
spectrum to be scale--invariant is independent of the value of $\beta$
(subject to the potential being able to drive inflation). This is interesting 
because previous models of inflation driven by exponential potentials 
have generated spectra that deviate significantly 
from the Harrison--Zeldovich form unless $\beta$ is sufficiently small. 
Indeed, models of steep inflation driven by 
such a potential in the RSII scenario 
predict $n_S \approx  0.944$ (for the case of 70 e--folds) and 
such a small value appears to be disfavoured by the recent 
WMAP data \cite{steep,wmap}. 
However, the above discussion indicates that it may be possible 
to realise inflation with a steep exponential potential, 
where the density perturbation spectrum is pushed close to  
scale invariance by the effects of the Gauss--Bonnet contribution. We  
perform a more detailed analysis of this possibility in the following 
Section. 

%%%%%%%%%%%%%%%%%%%%%%%%%%%%%%%%%%%%%%%%%%%%%%%%%%%%%%%%%%%%%%%
\section{Inflation driven by an exponential potential}
%%%%%%%%%%%%%%%%%%%%%%%%%%%%%%%%%%%%%%%%%%%%%%%%%%%%%%%%%%%%%%%
%\setcounter{equation}{0}
%\def\theequation{\thesection.\arabic{equation}}

Any successful model of inflation must satisfy the 
three key constraints 
that (a) sufficient inflation occurred to solve the horizon problem; 
(b) the amplitude of density perturbations is consistent with the  
COBE normalization of the CMB power spectrum and; 
(c) the spectral index must be sufficiently close to unity.
We now deduce the region of parameter 
space consistent with these constraints for a steep inflationary 
model driven by an exponential potential (\ref{exppot}). 

In determining the region of parameter 
space that is consistent with the observational constraints, 
it proves convenient to parametrize observable quantities in terms 
of the variable $x$. It follows from Eqs.~(\ref{Vsinhx}) 
and (\ref{exppot}) that
\begin{equation}
\label{phix}
\phi = \frac{1}{\beta\kappa_4} \ln \left[ \left(
\frac{\lambda b^{1/3}}{3\alpha V_0^2 \kappa_4^2}\right)^{1/2} \sinh x
\right] \,, 
\end{equation}
and the slow--roll parameters and density perturbation amplitude are 
evaluated from Eqs.~(\ref{epsilon}), (\ref{eta}), (\ref{xi}) 
and (\ref{scalar}), 
respectively: 
\begin{eqnarray}
\label{epsilonexp}
\epsilon &=&\frac{4 b^{1/2}}{27} \beta^2 \left( 
3 \alpha \lambda \kappa_4^2 \right)^{1/2} 
\frac{\sinh (2x/3) \, \sinh x \, \tanh x}{[b^{1/3} \cosh (2x/3) -1]^2} \,,
\\
\label{etaexp}
\eta &=& \frac{4 b^{1/6}}{9} ~\beta^2 \left( 3 \alpha \lambda \kappa_4^2 
\right)^{1/2} \frac{\sinh x}{b^{1/3} \cosh (2x/3) -1} \,, \\
\label{scalarexp}
A_S^2 &=& \frac{27}{1600\pi^2 b^{1/3}} ~\frac{1}{\beta^2 \alpha^2 \lambda}~
\frac{\left[ b^{1/3} \cosh (2x/3) -1 \right]^3}{\sinh^2 x} \,.
\end{eqnarray}
The slow--roll parameter, $\xi^2$, turns out to be simply $\xi^2 = \eta^2$ in
the case of a pure exponential potential.
The spectral index is then deduced implicitly by substituting 
Eqs.~(\ref{epsilonexp}) and (\ref{etaexp}) into Eq.~(\ref{scalartilt}).
The problem of developing a consistent model is now reduced to 
finding the value of $x$ that corresponds to 
the scale observable by COBE. Depending on the reheating temperature, 
this scale typically went beyond the Hubble radius some 
50--70 e--folds before the end of inflation. 

%%%%%%%%%%%%%%%%%%%%%%%%%%%%%%%%%%%
\subsection{Parameter space}
%%%%%%%%%%%%%%%%%%%%%%%%%%%%%%%%%%%
For an exponential potential, the value of the field, $x_N$, corresponding to 
$N$ e--folds before the end of inflation is given by Eq.~(\ref{Ngeneral}): 
\begin{equation}
\label{Nend}
N = -\left( \frac{27}{16 b^{1/3}} 
~\frac{\kappa_4^2}{\alpha \lambda}
\right)^{1/2}~
\frac{1}{(\beta \kappa_4)^2} \int^{x_{\rm end}}_{x_N} 
dx ~\frac{\left[ b^{1/3} \cosh (2x/3) -1 \right] \, 
\cosh x}{\sinh^2 x} \,,
\end{equation}
where we have substituted Eq.~(\ref{phix}).  
The integral in Eq.~(\ref{Nend}) can be evaluated analytically:
\begin{equation}
\label{Ndone}
N = -\left( \frac{27}{16 b^{1/3}}\right)^{1/2}~ \frac{1}{\left( 
\alpha \lambda \kappa_4^2 \right)^{1/2} \beta^2} ~
 \left[ f(x) \right]^{x_{\rm end}}_{x_N} \,,
\end{equation}
where we have defined the function $f(x)$: 
\begin{equation}
\label{deff}
f(x) \equiv \frac{2b^{1/3}}{\sqrt{3}} \tan^{-1} \left( 
\frac{2}{\sqrt{3}} \sinh (x/3) \right) + 
\frac{ 1-b^{1/3} \cosh (2x/3)}{\sinh x} \,.
\end{equation}
If $\alpha \Lambda \ll 1$, we can simplify the above
expression by specifying $b \approx 1$. In what follows, we will adopt this
simplification, which does not significantly alter the 
constraints on the parameters of the model. 
To constrain the parameter space $\{\alpha,\lambda,\Lambda \}$
we start by fixing the
value $x_N$ for a given number of e--folds, $N$, assumed to be
sufficient to give the flatness and homogeneity of the universe, and
the slope of the potential, $\beta$. By employing the COBE
normalization constraint (\ref{scalarexp}), 
we will extract the values of $\alpha$ and
$\lambda$ that satisfy this relation. 
The final step consists in obtaining
$\Lambda$ from Eq.~(\ref{tensiontune1}).

The end of inflation is calculated by noting that 
when $x \gg 1$ (i.e. when $q = 2/3$ in Eqs.~(\ref{epsilon2/3})), 
the slow--roll parameter $\epsilon$ 
simplifies to $\epsilon \propto V'^2/V^{5/3} \propto V^{1/3}$ 
and in this regime, $\epsilon$ {\em decreases} 
as the inflaton slowly rolls down its potential. More precisely, 
this implies that inflation can not end until  
$x$ is sufficiently small (i.e. $x \ll 1$) and it then follows from the 
discussion of Section 3.1 that the condition for 
inflation to end coincides with that of the 
RSII scenario (in the limit where $\rho \gg \lambda$). 
Consequently, taking the small $x$ limit 
of Eq.~(\ref{epsilon}) implies that
inflation ends when $2 \lambda V'^2 \approx 
\kappa_4^2V^3$, i.e., when $V_{\rm end} \approx 2\lambda \beta^2$ 
\cite{maartens,steep}\footnote{This further implies that 
the Gauss--Bonnet contribution does not influence the 
process of reheating after inflation has come to an end. 
Different mechanisms for reheating in the steep inflation scenario 
have been discussed previously \cite{steep,steepgw,steeppbh,steepcurv}.}.
Substituting Eq.~(\ref{Vsinhx}) then implies that 
\begin{equation}
\label{xend}
x_{\rm end} = 2 \beta^2 
\left( 3 \alpha \lambda \kappa_4^2 \right)^{1/2} \,,
\end{equation}
and substituting Eq.~(\ref{xend}) into Eq.~(\ref{Ndone}) with 
$f(x_{\rm end} \ll 1) = 2 x_{\rm end}/9$  
leads to an expression relating the value of $x$ at the end of inflation 
to its value $N$ e--folds from the end of 
inflation:
\begin{equation}
\label{xf}
x_{\rm end} = \frac{9}{2} \frac{f(x_N)}{N+1} \,.
\end{equation}
Hence, equating Eq.~(\ref{xend}) and Eq.~(\ref{xf}) implies that 
\begin{equation}
\label{al}
\alpha \lambda = \frac{81}{48}\frac{1}{\beta^4 \kappa_4^2} 
\left( \frac{f(x_N)}{N+1} \right)^2 \,.
\end{equation}

\begin{figure} \centerline{
\includegraphics[width=8.5cm]{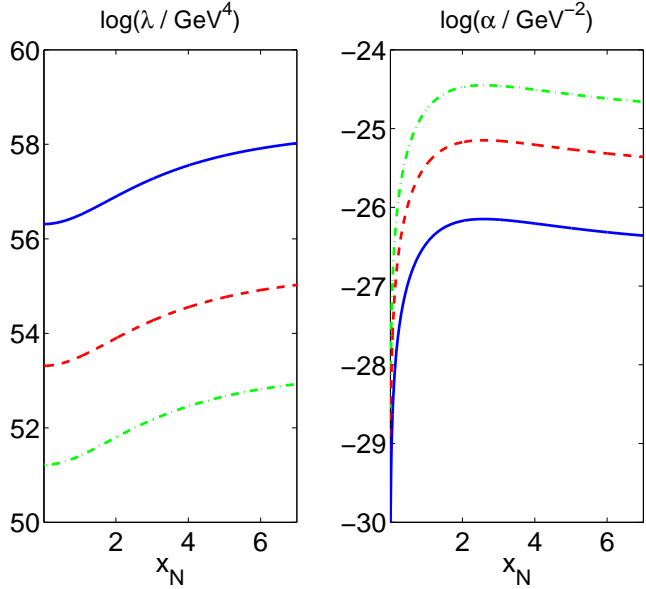} }%
\caption{\label{paramfig} Illustrating the relationships   
between the 
brane tension, $\lambda$, the Gauss--Bonnet coupling parameter, 
$\alpha$, and the value of $x_N$ that arise after 
COBE normalization of the scalar perturbation amplitude.  
The solid lines correspond to 
$\beta^2 = 10$, the dashed lines to $\beta^2 = 100$ and 
the dot--dashed lines to 
$\beta^2 = 500$. The value of $x_N$ was evaluated 
70 e--folds before the end of inflation and the constraints
are not significantly altered by considering lower values of $N$
such as $N=50$.}
\end{figure}

The COBE normalization then 
implies that Eq.~(\ref{scalarexp}) reduces to 
the constraint 
\begin{equation}
\label{a2l}
\alpha = \frac{10^{8}}{4 \pi^2} ~(\beta \kappa_4)^2
~\left(\frac{N+1}{f(x_N)} \right)^2 ~
\frac{\left[ \cosh (2x_N/3) -1 \right]^3}{\sinh^2 (x_N)} \,.
\end{equation}
For given values of $\{ \beta , N, x_N \}$, we may now
extract from Eqs.~(\ref{al})  and (\ref{a2l}) the values of the brane tension, 
$\lambda$, and Gauss--Bonnet coupling,
$\alpha$, that are consistent with the COBE normalization. Finally, 
Eq.~(\ref{tensiontune1}) fixes the five--dimensional
cosmological constant $\Lambda$. 
Fig.~\ref{paramfig} illustrates the dependence of $\lambda$ and  
$\alpha$ on $x_N$ for three different slopes of the potential, $\beta$.

%%%%%%%%%%%%%%%%%%%%%%%%%%%%%%%%%%%%%%%%%%%%%%%
\subsection{Predictions for the observables}
\begin{figure} \centerline{
\includegraphics[width=8.5cm]{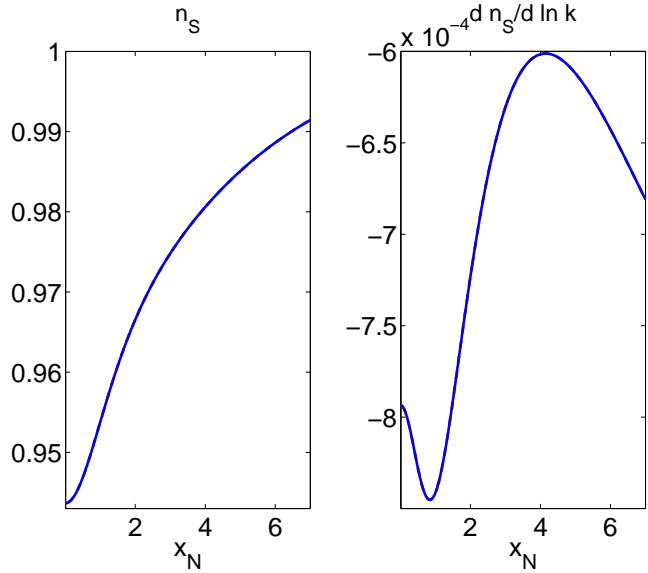} }%
\caption{\label{dndlnk} Illustrating the relationships   
between the spectral index, $n_S$, and its running, $d \ln n_S/d \ln k$,
and the value of $x_N$ that arise after 
COBE normalization of the scalar perturbation amplitude.  
The value of $x_N$ was evaluated 
70 e--folds before the end of inflation.}
\end{figure}

%%%%%%%%%%%%%%%%%%%%%%%%%%%%%%%%%%%%%%%%%%%%%%%
In determining the value of the spectral index, $n_S$, 
and its running, $dn_S/d\ln k$,
we observe from Eq.~(\ref{al}) that the combination,  
$\alpha \lambda \beta^4$, depends explicitly 
only on the value of $x_N$ for a given 
value of $N$. Moreover, this combination of parameters appears 
directly in the expressions (\ref{epsilonexp}) and (\ref{etaexp}) 
for the slow--roll parameters. 
It follows, therefore, that 
substituting Eq.~(\ref{al}) into Eqs.~(\ref{epsilonexp}) 
and (\ref{etaexp}) implies that the slow--roll 
parameters can be related directly to the value of $x_N$ {\em independently
of the slope of the potential} $\beta$: 
\begin{eqnarray}
\label{epsilonN}
\epsilon &=& \frac{1}{3} ~\frac{f(x_N)}{N+1} ~\frac{\sinh (2x_N/3) \, 
\sinh x_N \, \tanh x_N}{\left[ \cosh (2x_N/3) -1 \right]^2} \,, \\
\label{etaN}
\eta &=& \frac{f(x_N)}{N+1} ~\frac{\sinh x_N}{\cosh (2x_N/3) -1} \,.
\end{eqnarray}
Eqs.~(\ref{epsilonN}) and (\ref{etaN}) may then 
be substituted into Eqs.~(\ref{scalartilt}) and (\ref{runns}), 
thereby relating the 
spectral index and its running 
directly to $x_N$. Fig.~\ref{dndlnk} illustrates the
dependence of these two parameters on $x_N$. We observe that there 
is a lower limit of $n_S = 1 - 4/(N+1) = 0.944$, corresponding to 
the value predicted in the low energy (RSII) limit of steep inflation
(i.e. when $\alpha \rightarrow 0$) \cite{steep}. The spectral index
approaches unity if the $N$th
e--fold before the end of inflation occurred in the high energy limit
$x_N \gg 1$.
Similarly, the
expected value
for the running of the spectral index 
$d n_S/ d \ln k = -4/(N+1)^2 = -7.9 \times
10^{-4}$ is found in the limit $x_N \ll 1$ \cite{nunes}.

\begin{figure} \centerline{
\includegraphics[width=8.5cm]{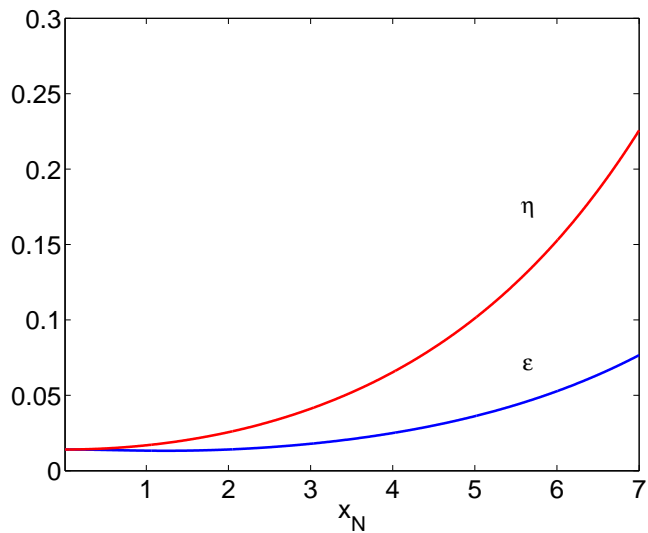} }%
\caption{\label{slowroll} Illustrating the variation of the slow--roll 
parameters, $\epsilon$ and $\eta$, on $x_N$.}
\end{figure}

Our approach thus far in this Section
has been to choose a value of $x_N$ for a given value of $N$, 
where the latter is chosen so that the horizon problem is automatically 
satisfied. However, there are two consistency checks that must  
be made to ensure that the above analysis is self--consistent. 
Firstly, one must ensure that inflation had indeed started by the time
$x = x_N$. In other words, we must verify that the slow roll parameters
are always less than unity in the range $x_{\rm end} < x<x_N$ 
for a chosen $x_N$. Fig.~\ref{slowroll} verifies 
that the $\epsilon$ and $\eta$ parameters indeed satisfy this
requirement. 

\begin{figure} \centerline{
\includegraphics[width=8.5cm]{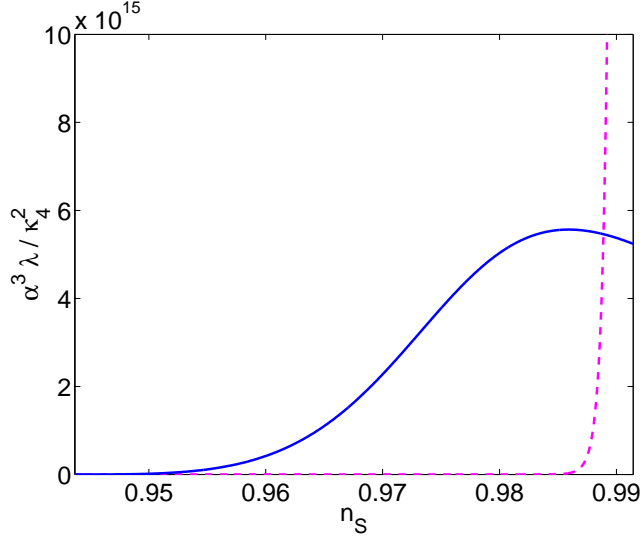} }%
\caption{\label{qgrav} Illustrating the relationship (solid line) 
between $\alpha^3 \lambda /\kappa_4^2$ and the spectral index, 
$n_S$, consistent with the COBE normalization constraint. The 
dashed line represents the critical case where the Planckian limit 
(\ref{upperlimit}) is just satisfied. The region of parameter 
space consistent with this constraint is located above and to the left 
of this line. This results in an upper limit on the allowed value of the 
spectral index of $n_S \approx 0.989$.}
\end{figure}

Secondly, 
in the present scenario, the 
assumption that the scalar field is confined to the brane 
becomes unreliable if the energy density of the inflaton field exceeds the 
five--dimensional Planck scale. We must therefore impose the constraint 
$\rho < \kappa_5^{-8/3}$ and this leads to a lower limit on the allowed 
value of $\alpha^3 \lambda$ for a given $x_N$: 
\begin{equation}
\label{upperlimit}
%x_N < \sinh^{-1} \left[ \left( \frac{\alpha^3 \lambda}{48 \kappa_4^2} 
%\right)^{1/6} \right] \,.
\alpha^3 \lambda > 48 \kappa_4^2 \sinh^6 x_N \,.
\end{equation}
The constraint (\ref{upperlimit}) effectively results in
an upper limit on the allowed value of the spectral index, as follows 
from Fig.~\ref{qgrav}. This limit can be quantified by 
noting that $\alpha^3
\lambda$ can be related directly to $x_N$ independently of the 
value of $\beta$ by 
combining Eq.~(\ref{al}) with Eq.~(\ref{a2l}). It follows that  
\begin{equation}
\label{a3l}
\alpha^3 \lambda = \frac{27 \times 10^{16}}{256 \pi^4} ~\kappa_4^2 ~
\left( \frac{N+1}{f(x_N)}\right)^2 ~
\frac{\left[ \cosh (2x_N/3) -1 \right]^6}{\sinh^4 (x_N)} \,,
\end{equation}
and, since Fig.~\ref{dndlnk} implies that the correspondence 
between $x_N$ and $n_S$ is one--to--one, 
we may relate $\alpha^3 \lambda/\kappa_4^2$ directly 
to $n_S$. The correspondence is shown in 
Fig.~\ref{qgrav}. 
Similarly, we may infer the region of parameter space consistent 
with the Planck limit (\ref{upperlimit}) and this is 
illustrated by the dashed line in Fig.~\ref{qgrav}. 
We verify that the constraint $\rho < \kappa_5^{-8/3}$ 
breaks down for $n_S \ge 0.989$ or, equivalently, for $x_N \ge 6$. 
Thus, we predict that for this model, the allowed values of 
the spectral index and 
its running  
are bounded both from above and below. For example,
in the specific case where $N=70$, we conclude that 
\begin{equation}
\label{bound}
0.944 \le n_S \le 0.989 \,,
\end{equation}
\begin{equation}
-0.85 \times 10^{-3} \le \frac{d n_S}{d \ln k} \le -0.60 \times
10^{-3} \,.
\end{equation}

These results {\it are
not} significantly altered by reducing the total number 
of e--folds to $N = 50$. The sensitivity of the constraints on 
the number of e--folds before the end of inflation is summarized in 
Table~\ref{tabela}.

\begin{table}
%\begin{ruledtabular}
\begin{center}
\begin{tabular}{|c|c|c|}
\hline
{~$N$~}& {\bf $n_S$ }& {\bf $- d n_S/d \ln k \times 10^{3}$} 
\\ \hline
10 &  ~0.636 -- 0.915~ & 25.0 -- 35.2 \\
20 &  ~0.809 -- 0.958~ & 6.87 -- 9.66 \\
30 &  ~0.871 -- 0.972~ & 3.15 -- 4.43 \\
40 &  ~0.902 -- 0.980~ & 1.80 -- 2.53 \\
50 &  ~0.922 -- 0.984~ & 1.16 -- 1.63 \\
60 &  ~0.934 -- 0.987~ & 0.81 -- 1.14 \\
70 &  ~0.944 -- 0.989~ & 0.60 -- 0.85 \\
\hline
\end{tabular}
\caption{\label{tabela}The allowed values of the spectral index, $n_S$,
and its running, $dn_S/d\ln k$, is weakly dependent on $N$, at least 
for $N\ge 40$.}
\end{center}
%\end{ruledtabular}
\end{table}

%%%%%%%%%%%%%%%%%%%%%%%%%%%%%%%%%%%%%%%%%%%%%%%%%%%%%%%%%%%%%%%%
\section{Summary and discussion}
%%%%%%%%%%%%%%%%%%%%%%%%%%%%%%%%%%%%%%%%%%%%%%%%%%%%%%%%%%%%%%%%
%\setcounter{equation}{0}
%\def\theequation{\thesection.\arabic{equation}}

In this paper, we have considered how the inclusion 
of a Gauss--Bonnet term in the bulk theory influences 
inflationary cosmology within the context of the Randall--Sundrum 
type II braneworld scenario. We have found that such a term can 
have a significant effect on the observational consequences of 
the scenario. In particular, we have focused on steep inflation, 
where the accelerated expansion of the braneworld is driven by an 
exponential potential with a logarithmic slope, $\beta \gg 1$. 
The effects of the Gauss--Bonnet contribution on the brane dynamics 
become significant 
at high energies and result in a density perturbation spectrum 
that can be very close to the scale--invariant (Harrison--Zeldovich) form. 
This is interesting given that steep exponential 
 potentials arise in a number of M--theory inspired models 
\cite{steepM}. 
Moreover, 
the numerical values of the spectral index and of its running are determined 
by the Gauss--Bonnet coupling parameter, $\alpha$, 
and the brane tension $\lambda$ (or, equivalently, the bulk 
cosmological constant, $\Lambda$) and are {\em independent} of 
the slope of the 
potential, $\beta$. 

We found that the value for the running of the spectral index is of
the order $d n_S/d \ln k \approx - 10^{-3}$.
Although preliminary analyses of the recent WMAP data 
have favoured somewhat smaller values \cite{wmap}, different 
authors have found little evidence for 
$d n_S /d \ln k
\neq 0$ \cite{seljak}. Nevertheless, 
improved observations will yield further information on the 
shape of the power spectrum and, consequently, on the validity of the 
model developed in the present work.

An important question that arises is how the spectrum 
of tensor (gravitational wave) perturbations is altered by the 
Gauss--Bonnet term. 
The calculation of the tensor perturbations in braneworld 
cosmology is more 
involved than that of the scalar perturbations, because the former 
extend into the bulk \cite{lmw}. The equation of motion for the tensor modes 
is derived from the linear perturbations of the bulk field equations, 
but the linearly perturbed junction conditions 
impose a boundary condition on the modes at the location of the brane. 
A detailed study of these equations is beyond the scope of the 
present work. However, the 
model we have considered predicts a potentially detectable signal 
of long--wavelength gravitational waves in the absence of 
the Gauss--Bonnet term \cite{steep,steep1,steepgw} and it would be 
interesting to investigate whether such a prediction is sensitive 
to the Gauss--Bonnet contribution. There is also the related 
question of whether the relationship between the scalar and 
tensor spectra is altered. In general, the two spectra are not 
independent and are related through a ``consistency'' relation, 
where the ratio of the amplitudes of the tensor and scalar perturbations 
is uniquely determined by the spectral index of the tensor spectrum. 
Such a relation represents a potentially observable signature of 
single-field inflationary models \cite{llkcba}.
It was recently shown that in a number of braneworld models, the 
consistency equation takes precisely the same form as that of 
standard inflationary cosmology \cite{steep1,consistent}. 
This raises the question of whether the Gauss--Bonnet contribution 
is able to lift the degeneracy.  

Finally, we conclude by highlighting a further consequence of 
the Gauss--Bonnet contribution on braneworld inflation. In Section 3.1, it 
was shown that at sufficiently high energies, the condition for 
inflation is that the pressure of matter on the brane should be negative, 
$p< 0$. This 
conclusion may have implications for cosmological models 
dominated by a tachyon field.
Following the work of Sen in understanding 
the role of the tachyon condensate in string theory \cite{sen}, there has been 
considerable interest recently 
in developing models of inflation driven by such a field, and  
a recent discussion of the prospects and 
problems associated with tachyon cosmology was presented by 
Gibbons \cite{gibbons}. The dynamics of a time--dependent and 
homogeneous tachyon field, $T$, may be described by the 
effective lagrangian ${\cal{L}} = -U (T) [1-\dot{T}^2]^{1/2}$, where 
$U(T)$ represents the tachyon potential. This implies that the 
pressure of such a field is given by 
\begin{equation}
p=-U(T) \sqrt{1-\dot{T}^2} \,,
\end{equation}
and is {\em negative--definite} for a positive--definite potential, $U>0$. 
Thus, if we consider the tachyon as a degree of freedom on the brane, 
the above discussion indicates that a Gauss--Bonnet contribution 
may allow inflation to proceed at sufficiently high energies 
with only a very weak dependence on the functional form of the 
tachyon potential. It would be interesting to explore this 
possibility further.   

\vspace{1cm}

\section*{Acknowledgements}
We thank E. Gravanis and N. E. Mavromatos for discussions and 
C. Charmousis for a helpful communication.
JEL is supported by the Royal Society. NJN is supported 
by PPARC.

\end{document}